\documentclass[reprint,
showpacs,preprintnumbers,showkeys,nofootinbib, amsmath,amssymb,
 aps,
prd,
]{revtex4-1}

\usepackage{amssymb}
\usepackage{amsmath}
\usepackage{graphicx}
\usepackage{color}
\usepackage{soul}

\def\beq{\begin{eqnarray}}
\def\eeq{\end{eqnarray}}

\renewcommand{\vec}[1]{{\bf #1}}

\def\Box{\square}                       

\def\na{\nabla}                         
\def\pa{\partial}                       

\def\={\ =\ }

\def\al{\alpha}
\def\be{\beta}

\def\de{\delta}

\def\la{\lambda}

\def\si{\sigma}
\def\om{\omega}
\def\ph{\varphi}

\def\Ga{\Gamma}
\def\De{\Delta}
\def\La{\Lambda}

\def\Om{\Omega}

\begin{document}

\preprint{PrePrint}

\title{Tensor instabilities at the end of the $\La$CDM universe}

\author{Giulia Cusin}
\email{giulia.cusin@unige.ch}
\affiliation{D\'epartement de Physique Th\'eorique and Center for Astroparticle
Physics, Universit\'e de Gen\`eve,
24 quai Ansermet, CH–1211 Gen\'eve 4, Switzerland}

\author{Filipe de O. Salles}
\email{fsalles@fisica.ufjf.br}
\author{Ilya L. Shapiro}
\email{shapiro@fisica.ufjf.br}
\altaffiliation{Also at Tomsk State
Pedagogical University, Tomsk, Russia.}
\affiliation{Departamento de F\'{\i}sica, ICE,
Universidade Federal de Juiz de Fora, 36036-330, MG, Brazil}

\date{\today}

\begin{abstract}
The unphysical spin-2 massive degrees of freedom in higher derivative
gravity may be either massive unphysical ghosts or tachyonic ghosts. In
the last case there is no Planck-scale threshold protecting vacuum
cosmological solutions from instabilities. Within the anomaly-induced
action formalism the photon-driven IR running of the coefficient of
the Weyl-squared term makes the ghost eventually becoming tachyon,
that should produce a gravitational explosion of vacuum. This effect
is stable under higher loop corrections and takes place also in known
versions of perturbative quantum gravity. However, the contribution
of massless fields in the far IR are not the same in flat and
de~Sitter spaces. In the asymptotically de~Sitter case one can observe
a kind of IR decoupling, which protects the cosmological solution from
the future tachyonic instabilities.
\end{abstract}



\maketitle

\section{\label{s1}Introduction}

It is well-known that fourth derivative terms are necessary in
the gravitational action, to render the theory consistent at the
semiclassical level (see \cite{birdav,book} for an introduction
to the subject and also \cite{PoImpo} for a recent pedagogical
review). In Quantum Gravity (QG) the same fourth derivative
terms provide renormalizability of the theory \cite{Stelle-77}.
At the same time, adding fourth derivative terms to the standard
Einstein-Hilbert action has an undesirable effect, since the
spectrum of the theory in the spin-2 sector gains a massive
unphysical ghost in addition to the usual healthy graviton. An
attempt to remove these unphysical states from the initial quantum
state leads to a non-unitary quantum theory. In the semiclassical
theory, when gravity is a classical background for quantum matter
fields, there is no problem with the unitarity of the gravitational
$S$-matrix, hence the condition of consistency can be reduced to
the requirement of stability of the physically relevant classical
solutions with respect to small metric perturbations. In the
present work we will discuss a new aspect of this problem,
related to the difference between the massive unphysical ghost
and the tachyon.

Let us start by fixing the notations. The action of gravity
corresponding to the renormalizable semiclassical
theory (for introduction one can see, e.g., \cite{birdav,book,PoImpo},
further references  therein and also recent more formal work
\cite{Ren-Curved})
consists of the following terms:
\beq
S_{vac}\,=\,S_{EH}\,+\,S_{HD}\,,
\label{vac}
\eeq
where $\,S_{EH}\,$ is the Einstein-Hilbert action with a
cosmological constant term,
\beq
S_{EH}
&=&
-\frac{1}{16\pi G}
\int d^4x \,\sqrt{-g}\,
\left\{R + 2\La \right\}
\label{S_EH}
\eeq
and $\,S_{HD}\,$ includes a minimal necessary
set of higher derivative terms,
\beq
S_{HD} = \int d^4x \sqrt{-g}
\{a_1 C^2 + a_2 E_4 + a_3 {\Box}R
+ a_4 R^2\}.
\label{S_HD}
\eeq
In this expression
\beq
E_4 = R_{\mu\nu\al\be}^2 - 4 R_{\al\be}^2 + R^2
\label{E}
\eeq
is the Gauss-Bonnet term (Euler density in $d=4$) and $C^2$ is the
square of the Weyl tensor. Furthermore, we use units corresponding to
$c=\hbar=1$ and denote with $\,M_p = 1/\sqrt{G}\,$
the Planck mass. The metric has signature $(+, -, -, -)$. For the
sake of simplicity we assume that the absolute value of $a_1$ is of
order one.

A general observation about the form of the action and
especially higher derivative terms (\ref{S_HD}) is in order.
One can deal with these terms in two different (albeit
physically equivalent) ways. First, it is possible to consider
the classical action of gravity being free of higher derivative
terms. Then one can observe that (different from the flat-space
QED, for instance) there are higher derivative divergences, corresponding to the running of these terms and hence the
non-local structures such as we shall discuss later on in
Eq. (\ref{W2}). This means that even if the terms (\ref{S_HD})
are not introduced at the classical level, the corresponding
quantum contributions emerge anyway. Second, one can follow
the standard QFT approach and introduce into the classical
action all those terms which will emerge with divergent
coefficients in the loop corrections. The main advantage of
this procedure is technical simplicity, because one can always
deal with a usual renormalization procedure, consider the
renormalization group running in a useful conventional way
\cite{book} etc. The last
observation is that the list of the terms in (\ref{S_EH})
and (\ref{S_HD}) follows from the fact that only these
structures emerge in the semiclassical approach at all
loop orders. This fact is known starting from the classical
work of Utiyama and DeWitt \cite{UDW}, and can be easily seen
from the covariant renormalizability and power counting
for the semiclassical theory \cite{Ren-Curved}.

The spectrum of the theory (\ref{vac}) linearized around a Minkowski
background contains in the spin-2 sector the massless graviton and
an additional spin-2 particle with mass of order $M_p$. If $a_1$ is
negative this particle has positive mass square, negative kinetic
energy and hence is a ghost. For a positive $a_1$ the mass square
is negative and the particle is a ghost and a tachyon at the same
time \cite{Stelle78}. We postpone the detailed description of these
two types of particles to the next section and now only make some
general observations.

According to the recent discussion in \cite{HD-Stab}, at least on a
cosmological FRW background massive ghost excitations are not generated
far below the Planck threshold of frequencies. The same result was found
much earlier \cite{star83,wave,HHR,GW-Stab} for de\,Sitter background.
Qualitatively similar considerations can be found in \cite{CreNic} for
generic models of massive gravity, where one can also meet massive
ghosts or tachyons. The result of \cite{HD-Stab} means that a massive
unphysical ghost can be destructive for the theory, but only if it
appears as a real physical particle. On the other side, the presence
of a ghost in the mass spectrum of the theory may not affect stability
if such a ghost is just a virtual particle in the vacuum state of the
theory.
One can note that the problem of massive ghost is a tree-level problem
of quantum theory. It is equivalent to the absence of stability in the
classical theory. In practical terms the conclusion of \cite{HD-Stab}
means that one can choose the initial condition in such a way that the
system does not develop instability for sufficiently long time.
According to \cite{HD-Stab} this choice of initial conditions
is possible for $a_1<0$.

On the contrary, for the case $a_1>0$ the extra massive particle is a
ghost and tachyon at the same time and satisfies an anti-oscillator
equation with exponential solutions. Then instabilities are unavoidable,
even without the presence of an external force. Moreover, the intensity
of the instability may be even enhanced if the mass of the tachyon
increases. There is no possible choice of initial conditions which
can help to control the instabilities in this case.

One can naturally ask whether we can define $a_1$ to be zero and
thus avoid to have both ghost and tachyon states in the spectrum.
The answer to this question is negative, because at the quantum
level the parameter $a_1$ runs due to the change of the energy scale.
In this work we will assume that gravity is classical and we take
into account only quantum effects of matter fields. In the last
section we will also comment on the possible role of quantum
gravity within some of the existing approaches.

Let us stress that the running of $a_1$ is a pretty well-explored
theoretical phenomenon. In particular, the full expressions for
the ``physical'' $\be$-functions in the momentum subtraction
scheme of renormalization were calculated in \cite{apco} and
\cite{fervi}. At low energies one can observe quantum decoupling,
which is a close analogous of the Appelquist and Carazzone theorem
in QED \cite{AC}. This means that in the present-day universe,
when the typical energy scale of the background is of the order of
$H_0\sim 10^{-42}\,GeV$, all massive fields are too massive to
affect the running of $a_1$ in the IR. Hence, in the late universe
it is sufficient to take into account quantum effects produced
by the unique massless particle, which is the photon.
Starting from
Sect. III it will be done in the framework of anomaly-induced
effective action, which is the most appropriate formalism for
classically conformal fields.

It is well-known that the renormalization group in gravity meets
a hard problem in identifying the energy scale $\mu$ with some
physical quantity related to gravity (see, e.g., the discussion
of this issue in \cite{DCCR} and also the scale-setting procedure
suggested in \cite{babic,Hrvoje}). At the same time, all these
difficulties concern only quantum effects of massive fields,
while the case of photon is different. The point is that
electromagnetic fields are massless and possess local conformal
invariance. Thanks to this symmetry one can integrate the conformal
anomaly \cite{DDI-76,Duff-77,Duff-94} and arrive at the reliable
form of the semiclassical corrections to the classical action. The
anomaly-induced effective action \cite{rie} can be regarded as
a local version of renormalization group \cite{asta}.

The purpose of the present work is to use the anomaly-induced
action to explore the running of $a_1$ in the far IR due to loop
effects of virtual photons on a cosmological background. This
running predicts that at some point in the distant future the
parameter $a_1$ changes sign and becomes positive. At this
instant the tachyonic mode with mass parameter unbounded from
above $m_2 \sim 1/a_1$ will manifest an explosive growth of
gravitational waves within the full range of frequencies starting
from zero to the Planck scale. This means that in a very short
time interval metric perturbations break down the linear regime.

The paper is organized as follows. Sect. \ref{s2} mainly serves
pedagogical purposes and describes the classification of ghosts
and tachyons in free theories with actions of both second- and
fourth-order in derivatives. We also discuss of the difference
between ghosts and tachyons in the effective field theory approach.
In Sect. \ref{s3} we summarize the main results concerning the
anomaly-induced effective action of gravity and explore the
effective equations of motion on cosmological backgrounds for
radiation-, matter- and cosmological constant - dominated
solutions. Sect. \ref{s4} includes the final results for the
physical running of $a_1$ in the same three cases. In Sect.
\ref{s5} we specialize our study to the late de\,Sitter phase
of the evolution of the universe and estimate the time period
of time for the universe to accelerate until the instant when
$a_1$ changes sign from negative to positive. Finally, in
Sect. \ref{s6} we critically discuss the list of assumptions
which have been introduced in our analysis and draw our conclusions.

\section{\label{s2}Ghosts and tachyons}

In this section we briefly describe the classification of free
fields into normal ``healthy'' ones, ghosts and tachyons. For
our purpose it is sufficient to consider flat space-time.

\subsection{\label{s21}Second-order theories}

For the second-order theory the general action of a free field
$\,h(x)=h(t,{\vec r})\,$ is
\beq
S(h) &\,=\,&
\frac{s_1}{2} \int d^4x \,\big\{ \eta^{\mu\nu}\pa_\mu h \pa_\nu h
- s_2 m^2 h^2 \big\}
\nonumber
\\
&\,=\,&
\frac{s_1}{2} \int d^4x \,\big\{ {\dot h}^2 - (\na h)^2
- s_2 m^2 h^2 \big\}\,.
\label{act}
\eeq
Both $s_1$ and $s_2$ are sign factors which take values $\pm 1$
for different types of fields. In what follows we consider all
four combinations of these signs.

It proves useful
to perform the Fourier transform in the space variables,
\beq
h(t,{\vec r}) \,=\,
\frac{1}{(2\pi)^3} \int d^3k \,
e^{i{\vec k}\cdot{\vec r}}\,h(t,{\vec k})\,,
\label{Fur}
\eeq
and consider the dynamics of each component $\,h\equiv h(t,{\vec k})\,$
separately. It is easy to see that this dynamics is defined by
the action
\beq
S_{\vec k}(h) &\,=\,&
\frac{s_1}{2} \int dt \,\big\{ {\dot h}^2 - k^2 h
- s_2 m^2 h^2 \big\}
\nonumber
\\
&\,=\,&
\frac{s_1}{2} \int dt \,\big\{ {\dot h}^2
- m_k^2 h^2 \big\}\,,
\label{act*}
\eeq
where
\beq
{\vec k}^2 = {\vec k}\cdot{\vec k}\,,
\qquad
\mbox{and}
\qquad
m_k^2 = s_2 m^2 + {\vec k}^2 \,.
\label{kmass}
\eeq
The properties of the field are defined by the sign of $s_1$ and $s_2$.
The possible options can be classified as follows.
\vskip 2mm

\noindent
{\it (i)}
\ {\it Normal healthy field} \ corresponds to $s_1=s_2=1$.
The kinetic energy of the field is positive and the minimal
action can be achieved for a static configuration. Also, the
equation of motion is of the oscillatory type,
\beq
{\ddot h} + m_k^2 h \,=\,0\,,
\label{pp}
\eeq
with the usual periodic solution.
\vskip 2mm

\noindent
{\it (ii)}
\ {\it Tachyon} \ has $s_1=1$ and $s_2=-1$. The classical dynamics of
tachyons is described in the literature, e.g.,
\cite{Sudarshan,Terletsky}, hence we present just a very brief review.

For relatively small momenta $m_k^2 < 0$ in Eq.
(\ref{kmass}) and the equation of motion is
\beq
{\ddot h} - \om^2 h \,=\,0\,,
\qquad
\om^2 = \left|m_k^2\right|\,.
\label{pm}
\eeq
Then we have an anti-oscillatory equation with exponential solutions,
\beq
h\,=\,h_1 e^{\om t}\,+\,h_2 e^{-\om t}\,,
\qquad
\om^2 = \left|m_k^2\right|\,.
\label{om}
\eeq
If the particle moves faster than light the solution is of the
oscillatory kind, indicating that such a motion is ``natural'' for
this kind of particle \cite{Sudarshan,Terletsky}.

An additional observation concerning tachyons is in order.
For a field interacting with an external gravitational
background, it is not impossible to have a situation where
the same wave changes from being a normal healthy state to a
tachyonic one in different parts of the space-time manifold.
In principle, this situation may produce very strong effects
at both quantum \cite{vanzella-q} and classical \cite{vanzella-c}
levels. In the present paper we are considering a qualitatively
similar situation, where the massive component of the gravitational
wave changes its fundamental properties due to the change of the
energy scale or due to the time evolution in de\,Sitter vacuum.
\vskip 2mm

\noindent
{\it (iii)}
\ {\it Massive ghost} \
has $s_1=-1$ and $s_2=1$. It is not a tachyon, because $m_k^2\geq 0$.
In this case the kinetic energy of the field is negative, but one can
postulate zero variation of the action and arrive at the normal
oscillatory equation (\ref{pp}). A particle with negative kinetic
energy has the tendency to achieve a maximal speed, but a free particle
can not accelerate, for this would violate energy conservation. Hence
a free ghost does not produce any harm to the environment, being
isolated from it.
\
However, if we admit an interaction with healthy fields, the tendency
of a ghost is to accelerate and transmit positive energy difference to
these healthy fields \cite{Veltman-63}, e.g., in the form of quantum
emission of the corresponding particles. There are many detailed
discussions of the fundamental problems at both classical and quantum
levels which arise in the theories with ghosts (see, e.g.,
\cite{Simon-90} and \cite{Woodard-review}). In general, the problem
of ghosts and the consequent conflict between renormalizability and
unitarity is one of the most important for quantum gravity, and
hence attracted a lot of attention. The list
of references on the possible approaches to avoid the problem of
higher derivative ghosts can be found in Ref. \cite{HD-Stab}. In this
work we also suggested a new approach to dealing with ghosts, which
is based on the effective field theory ideas, but is technically
classical and very simple. There are strong indications that the
creation of ghost states from vacuum is strongly suppressed for a
relatively weak cosmological background of the late universe. The
consideration presented below is based on the same idea, and assumes
the same approach.
\vskip 2mm

\noindent
{\it (iv)}
\ {\it Tachyonic ghost} \
has $\,s_1=s_2=-1$. For relatively small ${\vec k}^2$ we have
$m_k^2 < 0$. The kinetic energy is negative and the derivation of
the equations of motion requires an additional definition similar
to the non-tachyonic ghost case. After that, one can notice that
the equation of motion is of the anti-oscillatory type, eq.
(\ref{pm}) and the solutions are exponential (\ref{om}).

\subsection{\label{s22} Fourth-order gravity at the linearized level}

In the fourth-order gravity (\ref{vac}) the equations for the
metric perturbations in flat space can be easily obtained from
the more general ones on cosmological backgrounds \cite{GW-Stab}
(see also Eq. (\ref{main}) in Sect. \ref{s5}),
\beq
\ddot{\ddot h} + 2{\vec k}^2{\ddot h}  + {\vec k}^4 h
\,-\,\frac{1}{32\pi G a_1}\, \big({\ddot h} + {\vec k}^2 h \big)
\,=\,0\,.
\label{hhhh}
\eeq
It proves useful to introduce a new notation
\beq
\frac{1}{32\pi G a_1} &=& -\,s_2 \,m^2\,,
\label{m2}
\eeq
where $s_2 = - {\rm sign}\,a_1$ and $m^2>0\,$.
Then one can recast Eq. (\ref{hhhh}) into the form
\beq
\Big( \frac{\pa^2}{\pa t^2} + {\vec k}^2 \Big)
\,\Big( \frac{\pa^2}{\pa t^2} + m_{k}^2 \Big)h\,=\,0\,,
\label{factor}
\eeq
where $m_k^2\,=\,{\vec k}^2+s_2m^2\,$.
The solutions of the last equation can be different, depending
on the sign of $a_1$ and hence of $s_2$. The general formula
for the frequencies is
\beq
\om_{1,2} &\approx&
\pm\,i\big({\vec k}^2\big)^{1/2}
\,\,\,
\mbox{and}
\,\,\,
\om_{3,4} \approx
\pm \big(-\,m_k^2\big)^{-1/2}\,.
\label{osc}
\eeq

For a negative $a_1$ there are only imaginary frequencies and hence
oscillator-type solutions. On the contrary, for a positive $a_1$
we have $s_2=-1$ and the roots $\om_{3,4}$ are real, since in this
case $-m_{k}^2>0$ for sufficiently small ${\vec k}^2$.
Indeed, the first couple of roots corresponds to the massless
graviton, and the second couple to an extra massive particle.
According to our classification, this particle is a ghost for
$a_1 < 0$ and, simultaneously, ghost and tachyon for $a_1 > 0$
(see \cite{Stelle78} for a detailed discussion).

\subsection{\label{s23}Ghost vs tachyon}

The main difference between ghosts and tachyons is that a ghost
may cause instabilities only when it couples to some healthy
fields or to the background, while with tachyons there is no such
a protection.

The situation with ghosts can be kept under control in the effective
field theory framework, as it was recently discussed in \cite{Burgess}.
In an effective field theory there may be an apparent ghost due to the
low-energy expansion in the powers of ${\cal E}/M$, where ${\cal E}$
is the energy and $M$ is cut-off. At the same time, there could be
no ghost in the underlying fundamental quantum theory. In this
situation one can safely use an effective description for energies
${\cal E} \ll M$ and do not pay much attention to the presence of
ghosts. The situation in gravity is similar, but with an important
difference that the ghost-free UV completion of the renormalizable
theory is not known. Let us note that string theory can not be
regarded as such an UV completion, since the $\,R_{\mu\nu}^2$-like
terms which are the source of ghosts are removed in string theory
by a special transformation of the background metric \cite{zwei}
and not by a low-energy expansion. On the other side,
there are strong indications that the gravitational theory with
massive ghosts can be free of instabilities at low energies, since
for ${\cal E} \ll M_P$ the ghost may be a virtual particle and
effectively it is not created from vacuum \cite{HD-Stab}
(including Erratum of this paper). The reason is that a weak
gravitational background is not
providing sufficient density of energy to generate ghost as a real
particle excitation. Only for the typical frequencies of the Planck
order of magnitude ghost becomes destructive, while at much lower
energies one does not need to worry about its presence in the spectrum.
By the end of the day the situation is very close to the one in effective
field theories \cite{Burgess}.

On the contrary, no low-energy protection can be expected in the theory
with tachyons, because they produce instabilities independently on their
interaction to normal particles or on the intensity of the background.
In other words, for tachyons the exponential behavior (\ref{om})
occurs at all frequencies, and not only above the Planck threshold
\cite{HD-Stab}. Therefore, the difference between ghosts and
tachyons is expected to be critical for the low-energy regimes.

\section{\label{s3}Anomaly-induced effective action in the
cosmological setting}

In what follows we will use the formalism of anomaly-induced
effective action of vacuum, so let us start by describing an
application of this approach to the late-epoch cosmology.

For a general theory including $N_s$ massless conformal scalars,
$N_f$ massless Dirac fermions and $N_v$ massless vectors, the
anomalous trace of the energy-momentum tensor is \cite{birdav}
\beq
\langle T^\mu_\mu \rangle
&=&
-\,\big( \be_1C^2
+ \be_2E_4
+ \be_3{\Box}R\big)\,,
\label{T}
\eeq
where
\beq
(4\pi)^2\,\be_1
 &=& \frac{1}{120}\,N_s + \frac{1}{20}\,N_f +
\frac{1}{10}\,N_v\,,
\nonumber
\\
(4\pi)^2\,\be_2 &=& -\,\frac{1}{360}\,N_s
- \frac{11}{360}\,N_f - \frac{31}{180}\,N_v\,,
\nonumber
\\
(4\pi)^2\,\be_3 &=& \frac{1}{180}\,N_s
+ \frac{1}{30}\,N_f - \frac{1}{10}\,N_v \,.
\label{ombc}
\eeq
It was  already mentioned in the Introduction that in the late
universe the loops of all massive fields decouple from gravity.
The reason is that the typical energy scale of
the universe is usually defined by the Hubble parameter, which
has the present-day order of magnitude
$H_0 \propto 10^{-42}\,GeV$. This is a very small value if
compared even to the lightest known particles, e.g., the
standard estimate for neutrino is
$m_\nu \propto 10^{-12}\,GeV$. Therefore,  according to
the existing results on the gravitational decoupling \cite{apco,fervi}
in the present-day (and certainly later) universe one need to take
into account only the contribution of  photons, which are massless.
Hence one has to set $N_s=N_f=0$ and $N_v=1$ in the
expressions for the $\be$-functions (\ref{ombc}).
Then these $\be$-functions (\ref{ombc}) boil down to
\beq
\be^{IR}_{1,2,3}
 &=&
\frac{1}{10\,(4\pi)^2}\,\Big(1,\,-\frac{31}{18},\,-1\Big)
\,\equiv\,(\,\om,\,b,\,c\,)\,.
\label{abc}
\eeq
In order to find the anomaly-induced action one has to solve
the equation
\beq
\frac{2}{\sqrt{-g}}g_{\mu\nu}
\frac{\de {\bar \Ga}_{ind}}{\de g_{\mu\nu}}
=
-\langle T_\mu^\mu \rangle
=
\om C^2 + bE_4 + c{\Box} R\,.
\label{mainequation}
\eeq
The solution of this equation has been originally found in \cite{rie}
and was discussed, e.g., in \cite{PoImpo,MaMo}. Let us present only
the final result for the covariant local form of the solution with two
auxiliary fields \cite{a,MaMo},
\beq
\bar{\Ga}_{ind}
 &=& S_c[g_{\mu\nu}]
+  \int d^4x \sqrt{-g}
\,\Big\{\frac12 \,\ph\De_4\ph - \frac12 \,\psi\De_ 4\psi
\nonumber
\\
&-& \frac{3c+2b}{36}\,R^2
+\frac12\,\ph\,\Big[\,\sqrt{-b}\,
\big(E_4 -\frac23\,{\Box}R\big)\,
\nonumber
\\
&-& \frac{1}{\sqrt{-b}}\,
\om C^2\,\Big]
+
\frac{\om}{2 \sqrt{-b}}\,\psi\,C^2 \,\Big\}\,,
\label{finaction}
\eeq
where
\beq
\De_4
&=&
\Box^2 \,+\, 2R^{\mu\nu}\na_\mu\na_\nu
\,-\, \frac23\,R\Box \,+\,\frac13\,R_{;\mu}\,\na^\mu
\label{Pan}
\eeq
is a covariant, self-adjoint, fourth-derivative,
conformal operator \cite{Paneitz}.

The relevance of the auxiliary scalars $\ph$ and $\psi$ is based
on the fact that the boundary conditions for these fields are
equivalent to the boundary conditions for the two Green functions
of the same operator $\De_4$ in the non-local covariant form of
$\Gamma_{ind}$.  The importance to have two auxiliary fields has
been addressed in Refs. \cite{a,balsan,MoVa,MaMo} and \cite{PoImpo},
where the discussion of the role of conformal functional
$S_c[g_{\mu\nu}]$ can be also found.

\begin{figure}
\includegraphics[height=2cm,width=8cm]{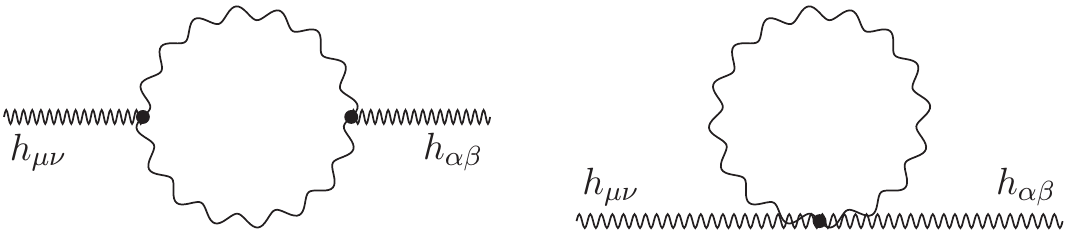}
\caption{Photon loops with two external gravitational lines.}
\label{Fig3}
\end{figure}

Let us make an important general observation. The dependence on the
conformal factor in the Weyl-squared term is exactly the same as the
one in the well-known logarithmic form factor for the massless quantum
fields\footnote{At the end of Sect. 6 we will discuss the
IR limits of this correspondence in the asymptotically dS space.}
\beq
\Ga_{Ws}
&=&
- \,\frac{\be_1}{2}\int \sqrt{-g}\,C_{\al\be\la\tau}
\,\log\Big(\frac{\Box}{\mu^2}\Big)\,C^{\al\be\la\tau}\,,
\label{W2}
\eeq
where $\Ga_{Ws}$ means $\Ga_{Weyl-squared}$ and
$\be_1=(160\pi^2)^{-2}$ for the photon. One has to remember
that the term (\ref{W2}) comes from quantum corrections (loop of
photon, in our case, see Fig. \ref{Fig3} for the illustration)
independently on whether the higher derivative term (\ref{S_HD})
is included or not into the classical vacuum action. Including
this term is useful to make the procedure of renormalization more
simple and regular. However, a finite term such as (\ref{W2})
will show up even if we set $a_1=0$. In
what follows it will be shown that an arbitrary $a_1$ will also
come from fixing the initial condition for the solutions of the
effective equations obtained by taking the term (\ref{finaction})
into account\footnote{One
see the details of this procedure in Eqs. (\ref{rad a1}),
(\ref{mat a1}) and (\ref{dS a1})}. The same happens in the case
of the quantum term (\ref{W2}), in the instant at which we fix the
value of $\mu$. Therefore it is not critically relevant for our
considerations that we started from the classical theory with
higher derivative terms (\ref{S_HD}), since the same terms emerge
from quantum corrections in any case. This consideration shows
also the deep relation between the anomaly-induced action and
renormalization group. This relation will be extensively
used in Sect. \ref{s6}.

Starting from (\ref{finaction}) one can derive the anomaly-induced
effective equations for various cosmological epochs. Consider the
effective action
\beq
\label{totalaction}
\Gamma
&=&
S_{vac}+\Gamma_{ind}\,,
\eeq
where the classical action $S_{vac}$ was defined in (\ref{vac}) and the
quantum correction $\Gamma_{ind}$ in (\ref{finaction}).

The equations for the auxiliary fields $\psi$ and $\phi$ are
\beq
\label{phi}
&& \sqrt{-g}
\left[\De_4 \phi + \frac{\sqrt{-b}}{2}\left(E_4-\frac{2}{3}
\Box R\right)-\frac{\omega}{2\sqrt{-b}} C^2 \right] = 0\,\,\,\,\,\,\,\,\,\,\,\,
\\
\label{psi}
&&
\sqrt{-g} \left[\Delta_4 \psi-\frac{\omega}{2\sqrt{-b}} C^2\right]
\,=\, 0\,.
\eeq
The conformal transformation of the metric
\beq
g_{\mu\nu}(x)
&=&
{\bar g}_{\mu\nu}(x)\,e^{2 \si(x)}\,,
\eeq
gives the following transformation for the quantities of our interest
(the reader can check \cite{Stud} for details)
\beq
&&\sqrt{-g} \,\Delta_4
\,=\,
\sqrt{- \bar{g}} \,{\bar \Delta}_4\,,
\quad
\label{icon2}
\nonumber
\\
&&
\sqrt{-g}\left(E_4-\frac{2}{3}\, \Box R\right)=
\sqrt{-\bar{g}}\left(\bar{E}_4
- \frac{2}{3}\, {\bar \Box} {\bar R}
+ 4 {\bar \De}_4 \si\right)\,.\,\,\,\,\,\,\,\,\,\,\,\,
\eeq
For the sake of simplicity we consider only the spatially flat metric,
\beq
g_{\mu\nu} \,=\,
\eta_{\mu\nu}\,a^2(\eta)
\,=\,
\eta_{\mu\nu}\,e^{2\si(\eta)}\,,
\eeq
where $\eta$ is conformal time, and use identities (\ref{icon2}).
Then Eqs. (\ref{phi}) and (\ref{psi}) become
\beq
\label{phif}
\Box^2 \phi
&=& -2\,\sqrt{-b}\,\Box^2 \sigma\,,
\\
\label{psif}
\Box^2 \psi
&=&
0\,,
\eeq
where the d'Alembertian is the flat-space one.

Solutions of Eqs. (\ref{phif}) and (\ref{psif}) can be found for
various epochs of the cosmological evolution of our universe, i.e.
separately during radiation, matter and during the de\,Sitter phase
which is the asymptotic future for the $\La$CDM universe.
\vskip 2mm

\noindent
{\large $\bullet$} \quad
During the radiation epoch, $a(\eta) \propto \eta$ and Eqs. (\ref{phif})
and (\ref{psif}) become
\beq
\phi^{(iv)}
 &=&
12\,\sqrt{-b}\,\eta^{-4}\,,
\\
\psi^{(iv)} &=& 0\,,
\eeq
where primes indicate derivatives with respect to conformal time.
The solutions have the form
\beq
\psi &=& d_0+d_1\,\eta+d_2\,\eta^2+d_3\,\eta^3\,,
\\
\phi &=& c_0+c_1\,\eta+c_2\,\eta^2+c_3\,\eta^3
\,-\,2\sqrt{-b}\,\log\Big(\frac{\eta}{\eta_0}\Big)\,,\,\,\,\,\,
\eeq
where $d_0$, \dots $d_3$,  $c_0$, \dots $c_3$, are integration
constants and $\eta_0$ is a generic reference time.
\vskip 2mm

\noindent
{\large $\bullet$} \quad
During the matter-dominated period, $a(\eta) \propto \eta^2$ and Eqs.
(\ref{phif}) and (\ref{psif}) become
\beq
\phi^{(iv)}  &=& 24\,\sqrt{-b}\,\eta^{-4}\,,
\\
\psi^{(iv)} &=& 0\,,
\eeq
with  solutions
\beq
\psi &=& d_0+d_1\,\eta+d_2\,\eta^2+d_3\,\eta^3\,,
\\
\phi &=& c_0+c_1\,\eta+c_2\,\eta^2+c_3\,\eta^3
\,-\,4\sqrt{-b}\,\log\Big(\frac{\eta}{\eta_0}\Big)\,,\,\,\,\,\,
\eeq
where $d_0$, \dots $d_3$,  $c_0$, \dots $c_3$, are integration constants
in general different from the corresponding ones in radiation.
\vskip 2mm

\noindent
{\large $\bullet$} \quad
During the de\,Sitter phase, we have $a(\eta) \propto |\eta|^{-1}$ and
Eqs. (\ref{phif}) and (\ref{psif}) become
\beq
\phi^{(iv)} &=& - 12\, \sqrt{-b}\,\eta^{-4}\,,
\\
\psi^{(iv)} &=& 0\,,
\eeq
with  solutions
\beq
\psi &=& d_0+d_1\,\eta+d_2\,\eta^2+d_3\,\eta^3\,,
\\
\phi &=& c_0+c_1\,\eta+c_2\,\eta^2+c_3\,\eta^3
\,+\,2\sqrt{-b}\,\log\Big(\frac{\eta}{\eta_0}\Big)\,,\,\,\,\,\,
\eeq
Here $d_0$, \dots $d_3$,  $c_0$, \dots $c_3$, are integration constants,
in general different from the corresponding ones in radiation- and
matter-dominated periods.

\section{\label{s4}Effective corrections to the classical action}

In Refs. \cite{wave,GW-Stab} and \cite{HD-Stab} tensor perturbations
of the classical theory (\ref{vac}) around a FRW background have been
investigated in details. The result is that the stability is fully
determined by the sign of the coefficient $a_1$ of the Weyl-squared term
and by the frequency of the perturbation $\,k=\left|{\vec k}\right|$. It
was found that

\begin{itemize}

\item If $a_1<0$ and  $k \ll M_{p}$\,, the theory  is stable under tensor perturbations,

\item If $a_1<0$ and above some frequency $k$, which is comparable to
$M_{p}$, the theory starts to be unstable under tensor perturbations,

\item If $a_1>0$\,, the theory is unstable under tensor perturbations
$\forall\, k$.
\end{itemize}

This result can be easily understood taking into account the physical
content of the higher derivative theory linearized around Minkowski
space, as it was explained in Sect. \ref{s2}. Our main interest is
the dynamics of tensor modes in the late universe. According to
\cite{HD-Stab}, for $a_1<0$ gravitational waves
start an exponential growth only if their frequencies are close to
the Planck order of magnitude. For this reason in what follows we
will restrict our analysis to frequencies satisfying $k \ll m_2$,
this means that the Planck scale physics is beyond our consideration.
One can assume that starting from the Planck scale the high energy
gravitational theory passes through some
qualitative change. For instance, in this regime the appropriate
description may be (super)string theory, which is free of the problems
of ghosts by construction (see subsection \ref{sub7}).

In the late universe the cosmological background is varying quite slowly
and one can define an effective coefficient of the Weyl-squared  term
(\ref{S_HD}), which takes into account semiclassical anomaly-induced
effects (\ref{finaction}),
\beq
a_1^{eff}(\eta)
&=& a_1 \,+\, \frac{\omega}{2\sqrt{-b}}\,
\big[\psi(\eta)-\phi(\eta)\big]\,.
\label{a1eff}
\eeq
Using the results of Sec. \ref{s3}, we obtain, in terms of physical time $t$,
\beq
\text{radiation}\hspace{1 cm}
a_1^{eff}&=&a_1^C+A_1\,t^{1/2}+A_2\, t+A_3\, t^{3/2}
\nonumber
\\
&+&\frac{\omega}{2}\log\left(\frac{t}{t_0}\right)\,,
\label{rad a1}
\\
\text{matter}\hspace{1 cm}
a_1^{eff}&=&a_1^C+A_1\,t^{1/3}+A_2\, t^{2/3}+A_3\, t
\nonumber
\\
&+&\frac{2\omega}{3}\log\left(\frac{t}{t_0}\right)\,,
\label{mat a1}
\\
\text{de\,Sitter}\hspace{1 cm}
a_1^{eff}&=&a_1^C+A_1\,e^{-Ht}+A_2\,e^{-2Ht}
\nonumber
\\
&+&A_3\, \,e^{-3Ht}+\om H t\,,
\label{dS a1}
\eeq
where $t_0$ is a reference time, the coefficients $A_i$ are arbitrary
integration constants and $a_1^C$ contains both the classical $a_1$
and the constant part of the semiclassical contributions. In each
epoch the $A_1$ and $a_1^C$ constant coefficients may be different
but for the sake of simplicity, we adopted the same notation for all
the cases.

The expressions (\ref{rad a1}), (\ref{mat a1}) and (\ref{dS a1}) come
from the anomaly-induced action (\ref{finaction}), but they can be
seen as a local version of the renormalization group running for the
parameter $a_1$. Since the quantum electromagnetic field is strictly
massless, one can trace this flow to the far IR. In the given theory
the UV turns out to be more complicated, because starting from some
high-energy scale massive fields start to contribute to the running
of $a_1$. The contribution of massive fields breaks down the elegant
form of the induced effective action, until the deep UV regime, when
all the fields can be treated as massless again \cite{Koma,Polchi}.

Let us additionally comment on the expressions (\ref{rad a1}),
(\ref{mat a1}) and (\ref{dS a1}). The complete solution for
$a_1^{eff}$ requires that the integration constants $a_1^C$
and $A_{1,2,3}$ are determined from experimental or observational
data and the solutions for different phases are connected by
requiring continuity.
However, it is difficult to put this program into practise,
because of the Planck suppression which makes impossible a direct
observation of the effects of $a_1$ or its running.  The unique way
to have an information about this coefficient is related to the
evidence that the present universe is stable under tensor
perturbations. We shall see in what follows that this is
sufficient to obtain some physically interesting information.

Our own existence shows that starting from the beginning of the
radiation epoch and along the cosmological evolution up to
present time, the
sign of $a_1^{eff}$ has been negative. If not, the tensor
perturbations would have been of the tachyonic type with energy
density of the Planck order of magnitude. Therefore, the proper
fact of our existence puts strong restrictions on the values of
$a_1^C$ and $A_{1,2,3}$ until the present time.

\section{\label{s5}Tensor perturbations in the asymptotically de Sitter phase}

According to the recent observations \cite{CMB,SNIa} our universe is
now entering into a phase of its evolution dominated by the
cosmological constant. The analysis of the
time-dependent parameter  $a_1^{eff}(t)$ during the last phase of the
evolution of the $\La$CDM universe deserves a special consideration
and will be explored in this section.

Let us parametrize tensor perturbations around the de\,Sitter
background ${\hat g}_{\mu\nu}$ as
\beq
g_{\mu\nu}
&=&
{\hat g}_{\mu\nu} + h_{\mu\nu}\,.
\eeq
Since the variation of $a_1$ in Eq. (\ref{dS a1}) is quite slow, it is
possible to consider a constant $a_1$ in the equations for the tensor
perturbations (see also discussion in \cite{GW-Stab} and \cite{HD-Stab}).

The equations of motion in the transverse and traceless (TT)
gauge can be cast into the form \cite{wave,GW-Stab,HD-Stab}
\beq
\label{dS}
\Big(\Box + \frac{4\Lambda}{3} + m_2^2 \Big)
\,\Big(\Box+\frac{2\La}{3}\Big)h_{\mu\nu} \,=\, 0\,,
\eeq
where $\Box$ is d'Alembertian operator on de\,Sitter background
and $m_2$ is the mass of the massive spin-2 mode
$\,m_2^2 = - (32\pi G\,a_1)^{-1}$.  Eq. (\ref{dS}) also agrees
with \cite{Gasperini,LuPo}. The theory propagates a massless
graviton $h_{\mu\nu}^{(m)}$, satisfying
\beq
\left(\Box+\frac{2\Lambda}{3}\right)h_{\mu\nu}^{(m)}=0
\eeq
and a massive spin-2 field $h_{\mu\nu}^{(M)}$, satisfying
\beq
\Big(\Box+\frac{4\Lambda}{3} + m^2_2\Big)
h_{\mu\nu}^{(M)}\,=\,0\,.
\eeq

Since we are interested only in the tensor part of the perturbation
$h_{\mu\nu}$, let us consider
\beq
\label{m}
h_{\mu\nu}
&=&
h_{\mu\nu}^{tens}
\,=\,
\left(
\begin{array}{cc}
0 & 0 \\
0 & h^{TT}_{ij}a^2
\end{array}
\right)\,,
\eeq
where we used TT-gauge
$\,\de^{ij}h^{TT}_{ij}=0$, $\,\pa^i h^{TT}_{ij}=0\,$
and set $\,a(t)=a(0)\,e^{Ht}$, with  $\,a(0)=1$.

For the analysis of tensor perturbations it is more practical to
use physical time $\,t$.
Furthermore, after we separate the $h^{TT}_{ij}$-mode, there is
no need to write indices and the field variable can be simply
denoted as $h=h(t)$.
Then the Eq. (\ref{dS}) can be written as
\beq
\label{main}
&&
\ddot{\ddot h} + 6 H \dddot{h}
+ \Big(11 H^2 + m_2^2 + \frac{2k^2}{a^2}\Big)\ddot{h}
\\
\nonumber
&& + \Big(6 H^3 + 3 H\,m_2^2
+ \frac{2 Hk^2}{a^2}\Big) \dot{h}
+ \Big(\frac{k^4}{a^4} + \frac{m_2^2\,k^2}{a^2}\Big)h
\,=\,0\,,
\eeq
where we have used the background constraint $\Lambda=3 H^2$ and
where dots indicate time derivatives. For $a_1=0$, Eq. (\ref{main})
reduces to the usual equation for General Relativity (GR)
tensor perturbations around
de\,Sitter background.

\begin{figure}
\includegraphics[height= 7cm,width=8.5cm]{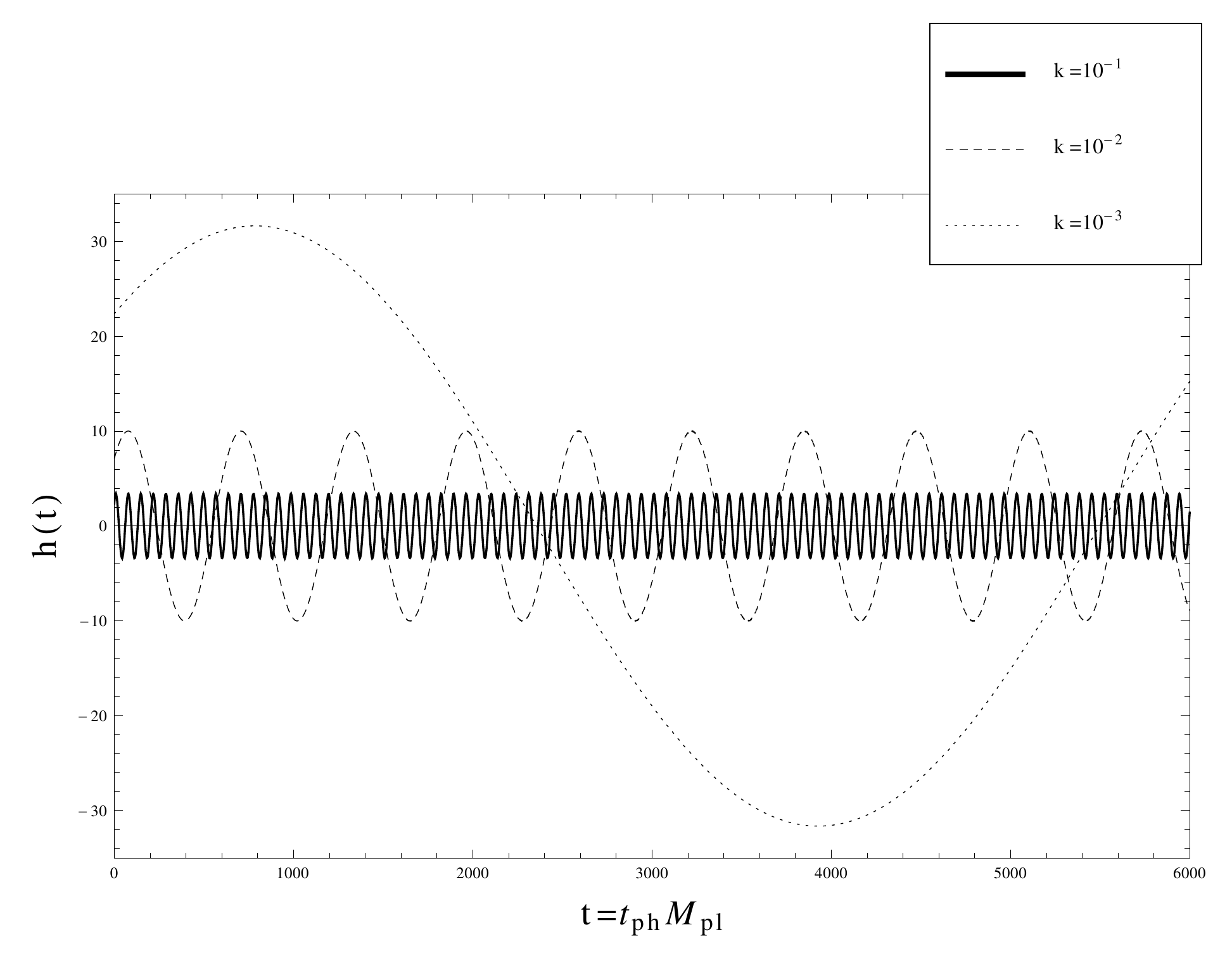}
\caption{Behavior of tensor perturbations in the case $a_1<0$, for
different frequencies. The frequency $k$ is measured in Planck units.}
\label{Fig1}
\end{figure}

\begin{figure}
\includegraphics[height= 6cm,width=8.5cm]{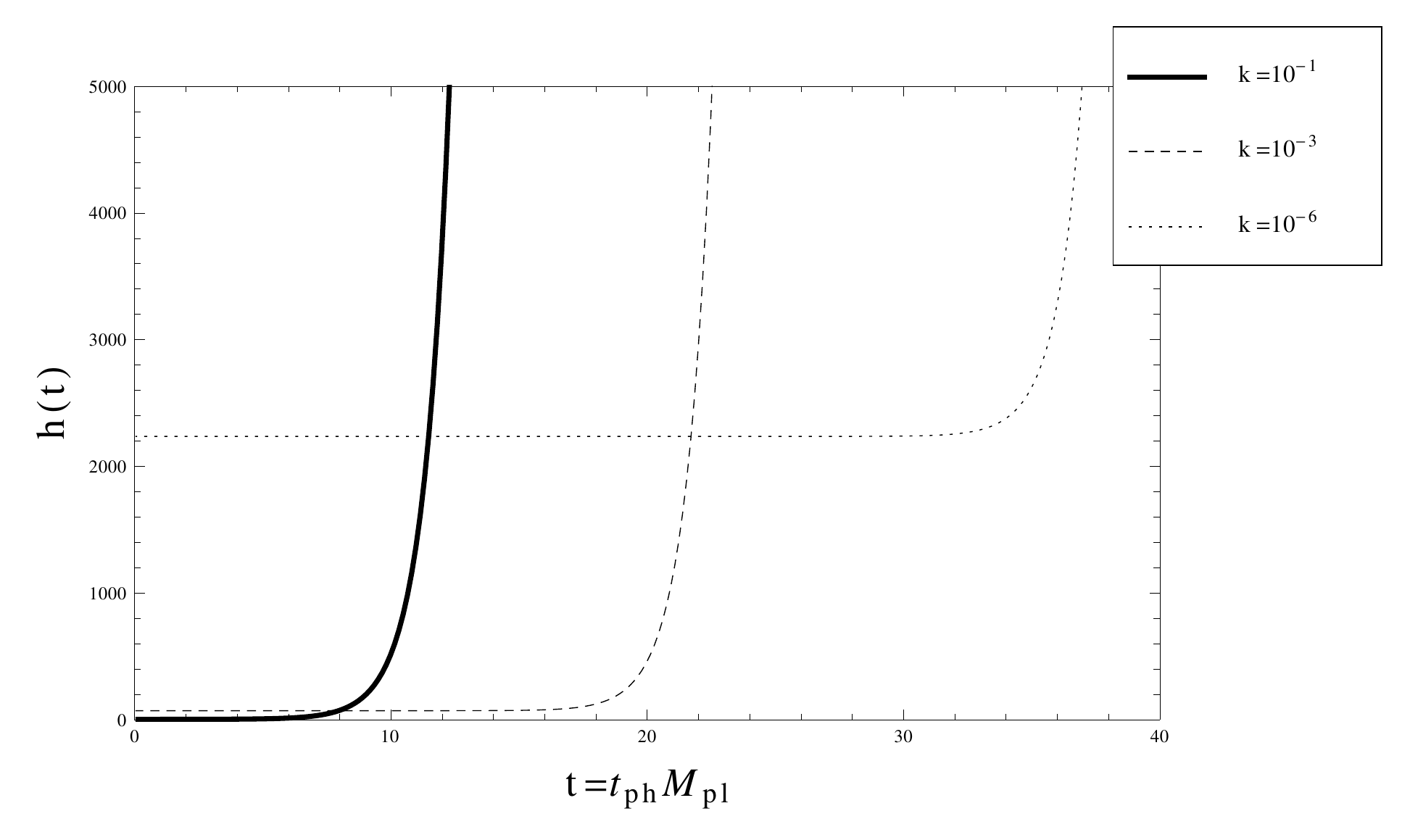}
\caption{Dynamics of tensor perturbations for $a_1>0$, for
three different frequencies $k = \left|{\vec k} \right|$. The
plot for the lowest frequency has been rescaled to fit the figure.}
\label{Fig2}
\end{figure}

The numerical investigation of Eq. (\ref{main}) for both signs of $a_1$
has been performed in \cite{HD-Stab}. As we have already mentioned, for
frequencies $k \ll M_p$ the case of $a_1<0$ shows oscillating solutions
without growing amplitudes. This behavior for three different
frequencies is shown in Fig. \ref{Fig1}. On the contrary, for $a_1>0$
one can observe a rapid growth of the metric perturbations for all
frequencies, see Fig \ref{Fig2}\footnote{Let us note that the numerical
integration can not be performed for positive $a_1$ which are too much
close to zero, since the growth of tensor perturbations is too fast in
this case.}

We want now to quantify which is the typical time scale at which
anomaly-induced corrections to the classical coefficient of the
Weyl square term, $a_1$, come into play. In pure de\,Sitter, we
found that the behavior of $a_1^{eff}$ is described by Eq. (\ref{dS a1}),
We focus our analysis on time-scales of order $t\sim 1/M_p$,
since this is the expected scale of the instabilities. On this
scales, the decaying exponentials in (\ref{dS a1}) vary slowly and
can be treated as constant functions. Hence, we can parametrize
Eq. (\ref{dS a1}) as
\beq
a_1^{eff}(t)
&=&
\tilde{a}_1 + \omega H\left(t-t_{dS}\right)\,,
\label{a1eff fin}
\eeq
where $t_{dS}$ is the time at which
the density of matter and radiation become effectively negligible
and a pure de\,Sitter phase starts.

Since we are not assisting an overproduction of gravitational
waves at present time, the nowadays value of $a_1^{eff}$ has to be
negative (i.e., we are in the regime in which tensor perturbations
are stable). As soon as $a_1^{eff}(t)$ crosses zero going to positive
values, the gravitational waves enter the unstable phase (the massive
ghost becomes tachyonic). Then an exponential instability in
the tensor sector instantaneously shows-up.

One can ask whether the stable phase, characterized by a
negative $\,a_1^{eff}$ will last for a long period of time.
We can assume that the stable phase in which
we are living today will last at least until the beginning of the
``pure'' de\,Sitter phase of the expansion of our universe, hence
we take  $\,\tilde{a}_1=a_1^{eff}(t_{dS})\,$ to be negative and of
order one. In the $\La$CDM model, with $\Om_{\La}=0.7$, the value
of the Hubble parameter in the pure de\,Sitter phase is
\beq
H_\La
&=&
\sqrt{\frac{\Lambda}{3}}
\,\simeq\, 0.8 H_0
\,\simeq\, 1.2\cdot10^{-42} GeV\,.
\eeq
Replacing $\,H=H_\La\,$ into (\ref{a1eff fin}) we find the
time $t_q$ at which tachyonic modes emerge in the spectrum of
gravitational waves,
\beq
t_q \,=\,
\frac{|{\tilde a}_1|}{H_\La\om}\,\simeq \,2.4\cdot 10^{13}\,yr\,=\,
2.4\cdot 10^{4}\,bi\,.
\label{tq}
\eeq
Hence with the assumption $|{\tilde a}_1|\sim 1$ the remaining
time until the gravitational wave explosion is about thousand
time longer than the time which already passed from the Big Bang.
Of course, since we have no experimental data to define the
present-day value of $|\tilde{a}_1|$, the estimate given above
has an ambiguity. According to Eq. (\ref{tq}) larger values of
$|\tilde{a}_1|$ correspond to a longer stable phase, while
smaller values of $|\tilde{a}_1|$ imply a shorter period of
stability before the tachyonic explosion.

Regardless of the quantitative side of this
prediction, it looks interesting that our knowledge of quantum
corrections to gravity is sufficient to know how the $\La$CDM
universe will end up. According to our analysis, there will be
an instant gravitational waves explosion due to the tachyonic
instabilities.

\section{\label{s6}Discussion of the assumptions}

We have started from a minimal set of hypotheses and arrived at
the conclusion that the final destiny of the $\La$CDM universe is
not a peaceful infinitely long slightly accelerated expansion in
the homogeneous and isotropic phase. On the contrary, the universe
will end up in a strong tachyonic explosion of tensor perturbations.
At first sight the intensity of this explosion
 (at least in the first order approximation) looks unrestricted,
since at the instant when $a_1^{eff}$ changes sign, the mass of
the unstable mode is infinite, according to Eq. (\ref{om}). Then
gravitational waves of all frequencies will experience a fast
exponential growth. However, since one can not trust the
semiclassical
approximation completely, the conservative estimate is that the
tachyonic modes will have, at most, energy density of the Planck
order of magnitude. Certainly, this is more than sufficient to
provide a dramatic effect on the geometry and matter contents
of the universe.

Another natural restriction comes from the fact that the instability
which we have found corresponds to linear perturbations. It might
happen that next orders in the perturbative expansion in $h_{\mu\nu}$
will restore the stability. However, from the practical side both these
restrictions do not matter too much, because even this ``restricted''
gravitational explosion should be capable to destroy the symmetry
(homogeneity and isotropy) of the metric, and thus lead to the strong
changes of the properties of space-time.

Since the result of our study looks so dramatic, it is interesting
to consider the list of all the approximations which have been
introduced in our analysis. Let us formulate various possibilities
to avoid the instabilities in a the form of questions and answers.

\subsection{\label{sub1}Completeness of the one-loop approximation}

\begin{figure}
\includegraphics[height=2cm,width=8.0cm]{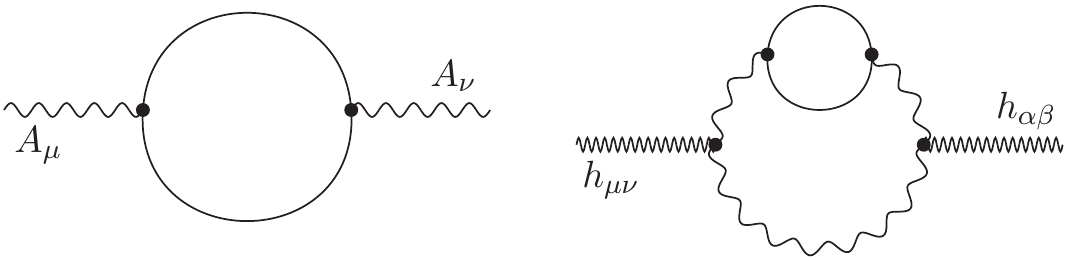}
\caption{One-loop diagram with electron loop and two external
electromagnetic lines and the diagram representing the
two-loop correction to the photon loop from Fig. \ref{Fig3},
with the electron loop insertion. The diagram at the left is
subject to decoupling in the IR.}
\label{Fig4}
\end{figure}

Can we expect some qualitative changes in the result by taking
higher-order loops into account? Formally, higher-loop
contributions do not change the sign of the $\be_1$-function
\cite{Koma,Polchi}, but this is not sufficient to draw the
conclusion about the relevance of higher-loop terms. In the UV
regime there will be higher logarithmic contributions compared
to the one-loop form factor (\ref{W2}) and this could produce
strong effect on the running of $a_1$. However, the situation
at low energies is quite different. Let us remember that
second- and higher-loop corrections to the one-photon bubble
include a loop of electrons or of other massive charged fermions,
as shown in Fig. \ref{Fig4}. Because of the Appelquist and Carazzone
decoupling theorem in QED, the contribution of such a loop is
suppressed at least by a factor
$\,\big(H_0/m_e)^2 \,\approx\, 10^{-77}$ for the dynamics of the
conformal factor. Since this is the part which governs the running
of $a_1^{eff}$, there are no chances that higher loops can
change the result found for the conformal factor dynamics.

At the same time, we have to remember that the goal of our study
is not the
dynamics of the conformal factor itself, but its interaction to
gravitational waves. Indeed, the term in the effective action which
we implicitly deal with is (\ref{W2}). At the two-loop order there
will be non-local corrections to the photon propagator and
photon-graviton vertex. Since we do not quantize gravity, the last
type of diagrams can be ruled out. The simplest diagram at higher
loops will be the one from Fig. \ref{Fig3} with the polarization
operator insertion shown in Fig. \ref{Fig4}, due to the electron loop.

Concerning this and further diagrams it is easy to see that the mass
which defines the scale of decoupling will be the mass of the quantum
fermion, while the typical energy of the external particle will be the
one of the gravitational wave. Finally, the relevant ratio which defines
the effective cut-off for the higher-loop contributions to the running
of $a_1$ is $\big({\cal E}_{GW}/m_e\big)^2$ for a tensor mode
with energy ${\cal E}_{GW}$. Of course this energy can be much
bigger than $\,H_0$. For instance, taking ${\cal E}_{GW}=1eV$ we have
``only'' ten-orders decoupling. Indeed, the gravitational wave
production with corresponding energy density will have a
very strong physical and geometrical effect.

It is worth noting that  we are dealing with the
diagrams which consist of the loops of matter fields and external
lines of gravity. Therefore, the one-loop part is given by the
contributions of free fields on a gravitational background.
For this reason there is no dependence on the coupling constants
at one loop order, and such dependence can be observed only
starting from the second loop \cite{birdav,book}.  Finally, the
one-loop effects of photons are {\it not} suppressed by mass, since
photon is massless. However, all particles to which photon is
coupled, are massive (and their masses are huge compared to
the gravitational waves energy) and that is why the two-loop
effects are indeed strongly suppressed.

The consideration presented above shows that the
one-loop results  (\ref{rad a1}), (\ref{mat a1}) and (\ref{dS a1})
are sufficiently robust in the framework of the semiclassical theory,
at least for gravitational waves with frequency smaller than the
electron mass. The case of waves with higher frequencies deserves
an additional investigation, but it can not change the qualitative
result which we obtained at the one-loop level.

\subsection{\label{sub2}The possible role of quantum gravity}

What about quantum gravity (QG) effects, which have been neglected
so far? It is not easy to give a definite answer due to the variety
of existing models of QG. Let us consider a short list of the
possibilities which have been better explored up to now\footnote{
Let us note that the problem of quantum stability of de\,Sitter
space and the validity of semiclassical approximation has been
extensively discussed in \cite{AnderMottola-2003} and more
recently in \cite{AnderMottola-2013} and \cite{AnderMottola-2014}.
The difference between our conclusions and the ones of these works
can be probably explained by the special role of tachyon tensor
modes which we consider in the present work.}.

\subsubsection{\label{sub3}Effective IR quantum gravity}

The standard effective framework for the IR effects of QG
assumes that General Relativity (GR) is a universal theory
of IR quantum gravity \cite{don94} (see also \cite{Iwa} for
earlier work in this direction and \cite{Polemic} for an
alternative opinion).

The first thing to note is that GR is not a
conformal theory, therefore a compact representation of quantum
corrections in the form (\ref{finaction}) is not possible. In this
situation one can make use of the renormalization group and of the
perturbative
$\be$-function $\be_1$ for the parameter $a_1$. The $\be_1$-function
gives, in the case of massless theory (e.g., quantum GR), the
coefficient in the logarithmic form factor (\ref{W2}). Therefore,
it can be still considered a good approximation in the present case.
In this fremework the change of sign of the renormalization group
equation (\ref{a1eff fin}) looks unclear, at least beyond the
one-loop level. The reason is that higher-loop contributions
of QG are beyond our control, because QG based on GR is
non-renormalizable. Still we can draw some conclusions about the
effective quantum gravity. At the one-loop level, the Weyl-squared
quantum counterterm is known to be gauge-fixing dependent \cite{KTT}.
Then one can provide any desirable value of $\be_1$ by a special
choice of the gauge-fixing parameters.  This also means that the
corresponding contribution vanishes on-shell and therefore can not be
regarded as a physical effect. Furthermore, similar gauge-fixing
dependence is expected for the leading logarithmic corrections at
higher loops. Therefore, while the subject is not completely clear,
one can suppose that the leading contributions of QG in this
framework will be sub-logarithmic. As we have already seen
in subsection \ref{sub1}, this means that the change of
sign of $\be_1$ in (\ref{ombc}) due to effective
low-energy QG is very unlikely.

\subsubsection{\label{sub4}Weyl conformal gravity}

In this case the form (\ref{finaction}) of quantum corrections is
available and the effect of QG is to increase the positive value of
$\be_1$. The reason is that the contribution of conformal QG is
positive \cite{frts82,amm92,Weyl}, exactly as the one of all matter
fields (\ref{ombc}). The expression for $N_v$ copies of massless
vector fields has the form
\beq
\be_{1}
&=&
\frac{1}{(4\pi)^2}\,\Big(\frac{N_v}{10} \,+\, \frac{199}{30}\Big)\,,
\label{extend-W}
\eeq
where the last term is the QG contribution.
It is easy to see that no cancellation is possible, and for $N_v=1$
the QG contribution shortens the period of time until the sign
transition for $a_1^{eff}$ of about one order of magnitude.

\subsubsection{\label{sub5}General version of QG with fourth derivatives}

In this case, exactly like in the effective QG described in
subsection \ref{sub3}, the form of quantum corrections of the type
(\ref{finaction}), is not viable, because the original theory is not
conformal, hence there is no anomaly to integrate. At the same
time the $\be$-function for the parameter $a_1$ is well-defined,
free of ambiguities \cite{frts82,avbar,a} and, according
to well-verified calculations \cite{frts82,avbar,Gauss} has the
same positive sign as (\ref{ombc}). The overall $\be_1$-function
for $N_v$ copies of massless vector fields has the form
\beq
\be_{1}
&=&
\frac{1}{(4\pi)^2}\,\Big(\frac{N_v}{10} \,+\, \frac{133}{10}\Big)\,.
\label{extend}
\eeq
In the physically interesting case we have one photon, $N_v=1$, and
no cancelation can be expected. Moreover, in the far IR the ghost
which provides some part of the QG contribution in (\ref{extend}),
is supposed to decouple, and we come back to the situation described
in subsection \ref{sub3}.

\subsubsection{\label{sub6}Superrenormalizable QG}

The superrenormalizable models of QG have been first formulated
in \cite{highderi} on the basis of a polynomial action and shortly
after that in \cite{Tomboulis-97} on the basis of an action of
gravity non-polynomial in derivatives. In the polynomial case this
model of QG has many (tensor and scalar) ghosts. In the non-local
model of \cite{Tomboulis-97} there are no ghosts at the tree-level,
however loop corrections lead to the emergence of an infinite
amount of massive ghost-like states, corresponding to complex poles
\cite{CountGhost}.

In both cases quantum corrections are well-defined, but the
$\be_1$-function for the parameter $a_1$ can be modified by
adjusting the terms of third- and fourth-orders in curvature
tensor \cite{CountGhost}. Therefore, it is not difficult to construct
the theory with the cancelation of the photon contribution in
(\ref{ombc}). However,
this cancelation will take place in the UV region, where the quantum
effects of ghosts are not suppressed. There is no systematic study
of decoupling of massive modes in higher derivative QG, but in
principle one can expect that in the IR the massive ghost states
will decouple and the low-energy situation will be covered by the
effective QG, as described in subsection \ref{sub3}. After all,
superrenormalizable models QG can not be expected to change the
sign of $\be_1$ in the IR limit.

\subsubsection{\label{sub7}String theory}

In (super)string theory the terms providing the $\be$-function for
the parameter $a_1$ are usually removed by means of the Zwiebach
transformation \cite{zwei,dere,tse}). The procedure is ambiguous
\cite{marot}, but, by construction the $\be$-function for the
parameter $a_1$ is zero. At the same time, string theory is not
supposed to provide a significant correction to the Quantum Field
Theory results at low and very low energies, for
otherwise we would observe such corrections in precision
experiments, e.g., the ones that test QED and Standard Model
calculations. Therefore using  string theory to evaluate the
$\be_1$-function in the far IR is not reasonable from a conceptual
point of view.
\vskip 2mm

In conclusion, we can see that QG can provide the change of
sign of the $\be$-function for $a_1$, but only at high energies
and in the framework of superrenormalizable models of QG of
\cite{highderi} and \cite{Tomboulis-97}. On the other hand,
the IR limit of the QG theory with a number of massive states
(ghosts and normal particles) should be taken with great care
\cite{don94,Polemic} and it is expected that the sign of the
$\be_1$-function in IR would remain positive. Hence, we arrive
at the conclusion that the main result concerning the positiveness
of the $\be$-function for the parameter $a_1$ remains robust even
if gravity is quantized.

\subsection{Completeness of the anomaly-induced approximation}

As we have seen, there is practically no way to avoid the change
of sign of $a_1$ within the anomaly-induced effective action
(\ref{finaction}). However, we know that this effective action is
based on the artificial Minimal Substraction ($\overline{\rm MS}$)
scheme of renormalization. This scheme is always working perfectly
well in the UV. On the other side, it is known that for the theory
of massive
quantum fields, this scheme is {\it not} working well in the IR,
because of the decoupling theorem \cite{AC}, which was also
obtained for semiclassical gravity in \cite{apco,fervi} (see also
\cite{PoImpo} for the review and further references). In the flat
space-time the masslessness guarantees that the main features of
the UV will repeat in the IR, due to the UV-IR ``duality'' of the
logarithmic form factor such as the one in (\ref{W2}). Is it
true that the situation is the same in curved space-time? The
question is not simple, especially in the asymptotically dS
space, which has a natural IR cut-off scale $H$.

For instance, the scalar curvature in the space with $\si(t)=H_0t$
(with $H_0=const$)
has the global scaling similar to the $\Box$ operator, namely
$\,R=-6e^{-2\si}\cdot \si''\,^2\,$ in the spacially
flat case, but the derivatives here are with respect to conformal
time $\eta$, where $\,dt=a(\eta)d\eta$. As a result $R$ does not
run with time, instead it is a constant\footnote{Authors are very
grateful to A.A. Starobinsky for suggesting this example as an
argument against IR running in de~Sitter space.}, $R=-12H_0^2$.
Similar consideration applies to the operator $\Box$ in the logarithm
of the form factor in (\ref{W2}). For instance, let us consider
$\Box$ acting on a scalar field $\ph$. For the sake of
simplicity we assume that this scalar depends only on time,
$\ph=\ph(t)$. Then a very simple calculation leads to the result
\beq
\Box \ph &=& {\ddot \ph} + 3 H_0 {\dot \ph}\,.
\label{box}
\eeq
One has to note that in the last equation the dependence $\ph(t)$
corresponds to the fiducial (flat) metric and, therefore, time
derivatives of $\ph$ are expected to be of the same order of
magnitude in different epochs. Thus, Eq. (\ref{box}) shows
that the dependence of time drops out from the coefficients
of the equation for perturbations. Similar calculation for the
Weyl-squared term provides the result
(here we kept the possibility to $H$ be non-constant for the
sake of generality)
\beq
&&\sqrt{-g}C^{\al\be\rho\tau}
\,\Box\,C^{\al\be\rho\tau}
=
\nonumber
\\
&=&\sqrt{-{\bar g}}{\bar C}^{\al\be\rho\tau}
\,\big\{
\pa_t^2 - 4H\pa_t - 2{\dot H} - 10H^2
\big\}
\,{\bar C}^{\al\be\rho\tau}
\nonumber
\\
&+&
8\sqrt{-{\bar g}}{\bar C}^{\al\be\rho\,t}
\,\big\{
H\pa_t - {\dot H} + 2H^2\big\}
\,{\bar C}^{\al\be\rho\,t}\,.
\label{Weyl-Box}
\eeq

The dependence of space coordinates does not change the result
qualitatively, if the wave vector is taken in the physical space,
that is after the rescaling $\,{\vec k} \to {\vec k} \cdot e^\si$.

The expression (\ref{Weyl-Box}) shows that in the
asymptotic future, where the background becomes very close to
de~Sitter space,  there will not be physical running
of $a_1$. The reason is that the extreme IR corresponds to
dS and has its own scale $H$, such that the correspondence
between logarithmic asymptotic (\ref{W2}) and anomaly-induced
action (\ref{finaction}) gets violated. Recently, explicit
calculations of the curvature-curvature correlators on de~Sitter
background have been performed in \cite{Ver}. It is remarkable
that these calculations (performed by completely different method)
have also shown the absence of leading-log corrections in
far IR. Our interpretation of this result is a specific IR
decoupling in the IR on de~Sitter space background, where
$H$ plays the role of the natural minimal mass scale.

Indeed, according to the same logic, the results for the
radiation- and dust-dominated epochs, Eqs. (\ref{rad a1})
and (\ref{mat a1}) are valid and the same is true for the
beginning of the transition period between matter and
cosmological constant dominating epochs. However, since we
are now in the situation when Hubble parameter is quite close
to the future constant value $H$, this means the universe
is entering the epoch of IR decoupling of all quantum effects
in the higher derivative sector of the theory, including the
ones of photons. Hence in the real situation the transition
to the positive sign of $a_1$ is very improbable.

\section{Conclusions}

The anomaly-induced effective action of vacuum is very efficient
tool for a local implementation of the renormalization group
running. According to this approach, the semiclassical effects
drive the coefficient $a_1$ of the Weyl-squared term in the vacuum
(gravitational) action to the positive side. It turns out that the
one-loop contribution is exact in this case, hence the future of
the universe should be related to the tachyonic instabilities.
Since the time-dependence of $a_1$ is linear, one can easily calculate
the time which remains until the change of its sign, that would be
an instant when the massive ghost transforms into tachyonic ghost.

The difference between the two types of ghosts is dramatic. For
the ``usual'' ghost there is a mass threshold and hence a possible
situation when the ghost is harmless. The situation can be
understood in the spirit of considerations presented in
\cite{Simon-90}. Let us note that he approach pursued in
these papers by Simon and Parker and Simon is effectively
equivalent to the one of \cite{HD-Stab}. In both cases Planck
frequencies are ``forbidden'' without clear explanation (we
believe that one can find a solution of this problem, but this
will be discussed elsewhere). Furthermore, in both cases a
physically reasonable
choice of initial conditions can be done, as a result the system
does not fall into the run-away solutions. As a result higher
derivative terms produce only tiny, Planck-suppressed corrections
and can be simply ignored. For the gravitational
waves this means that the dynamics of metric perturbations is
essentially the same as in GR, as it was actually shown by
Starobinsky in the second paper of \cite{star83}.

As far as ghost becomes also tachyon, no mass threshold exist
and no reasonable choice of initial conditions can be done.
Then an instant explosion of gravitational waves starts at all
frequencies, independent on initial conditions. Our calculations
show that in the framework of anomaly-induced effective action
the transition to tachyonic ghost is unavoidable in the far future
of the $\La$CDM universe.
It is easy to show that the higher-loops corrections do not change
this result, at least for the frequencies below the electron mass
scale.

However, in the dS-background case, which is a final stage of the
$\La$CDM universe, there is a qualitatively new type of IR
decoupling which takes place even for massless fields. As a
result, the anomaly-induced approximation, which is based on
the $\overline{\rm MS}$ scheme of renormalization becomes
non-appropriate in the IR, exactly as it happens in the massive
theory. Therefore, there are no real chances of the tachyonic
explosion of the universe even in the far future, since no
change of sign of $a_1$ can be expected.

\begin{acknowledgments}

The authors are grateful to M. Maggiore and A.A. Starobinsky for
stimulating discussions.
The work of G.C. is supported by the Swiss National Science Foundation.
F. Salles is grateful to FAPEMIG and CAPES for partial support.
I.Sh. is grateful to CNPq, FAPEMIG and ICTP
for partial support of his work and especially to the D\'epartement
de Physique Th\'eorique and Center for Astroparticle
Physics at the Universit\'e de Gen\`eve for partial support and
kind hospitality during his sabbatical period. Also, I.Sh. is
grateful to the Mainz Institute for Theoretical Physics (MITP)
and to the International Institute of Physics (IIP) in Natal
for their hospitality and partial support during the completion
of the first and last versions of this work.

\end{acknowledgments}


\end{document}